\documentclass[12pt]{article}
\usepackage{epsfig,amssymb,amsmath,multirow,}

\usepackage{graphicx,epsfig}

\usepackage{dcolumn}
\usepackage{bm}

\catcode`\@=11
\DeclareMathAlphabet{\mathpzc}{OT1}{pzc}{m}{it}

\font\cmss=cmss12 
\def\1{\hbox{{1}\kern-.25em\hbox{l}}}
\def\bfZ{\relax{\hbox{\cmss Z\kern-.4em Z}}}

\def \be  {\begin{equation}}
\def \ee  {\end{equation}}
\def \ba  {\begin{eqnarray}}
\def \ea  {\end{eqnarray}}
\def \baa {\begin{eqnarray*}}
\def \eaa {\end{eqnarray*}}
\def \bb  {\begin {thebibliography} }
\def \eb  {\end{thebibliography}}
\def \lab #1 {\label{#1}}

\def \matrix #1 {\left(\begin{array}{cc} #1 \end{array}\right)}

\newcommand{\as}{\ifmmode\alpha_{\rm s}\else{$\alpha_{\rm s}$}\fi}
\newcommand{\asbar}{\ifmmode\bar{\alpha}_{\rm s}\else{$\bar{\alpha}_{\rm s}$}\fi}

\newcommand{\Z}{{\mathbb Z}}

\font\cmss=cmss12 
\def\inbar{\,\vrule height1.5ex width.4pt depth0pt}
\def\IC{\relax\hbox{$\inbar\kern-.3em{\rm C}$}}
\def\IZ{\relax{\hbox{\cmss Z\kern-.4em Z}}}
\def\IR{{\hbox{{\rm I}\kern-.2em\hbox{\rm R}}}}

\def\IP{{\hbox{{\rm I}\kern-.2em\hbox{\rm P}}}}
\def\II{\hbox{{1}\kern-.25em\hbox{l}}}

\newbox\lett\newdimen\lheight\newdimen\lwidth
\def\ontop#1#2{\setbox\lett=\hbox{#2}\lheight\ht\lett
\multiply\lheight by 12 \divide\lheight by 10\relax%
\lwidth\wd\lett \multiply\lwidth by 8 \divide\lwidth by 10\relax #2\kern-
\lwidth%
\raise\lheight\hbox{{$\scriptstyle #1$}}\kern.1ex}

\def\XXint#1#2#3{{\setbox0=\hbox{$#1{#2#3}{\int}$}
     \vcenter{\hbox{$#2#3$}}\kern-.5\wd0}}

\begin{document}

\begin{titlepage}

\thispagestyle{empty}

\vspace*{1cm}

\centerline{\large \bf One String to Rule Them All: Neutrino Masses and Mixing Angles}

\vspace{1cm}

\centerline{\sc Jordan Gemmill, Evan Howington and Van E. Mayes}

\vspace{2mm}

\centerline{\it Department of Physical and Applied Sciences,}
\centerline{\it University of Houston-Clear Lake}
\centerline{\it Houston, TX 77058, USA}

\vspace{2cm}

\centerline{\bf Abstract}

\vspace{5mm}

The 
correct quark and charged lepton mass matrices along with a nearly correct CKM matrix 
may be naturally 
accommodated  
in a Pati-Salam model constructed from intersecting D6 branes 
on a $T^6/(\Z_2 \times \Z_2)$ orientifold. 
Furthermore, near-tribimaximal mixing for neutrinos may arise naturally due to the structure
of the Yukawa matrices.  Consistency with the quark  and charged lepton mass matrices 
in combination with obtaining near-tribimaximal mixing fixes the Dirac neutrino mass matrix
completely.  Then, applying the seesaw mechanism for different choices of right-handed neutrino
masses and running the obtained neutrino parameters down to the electroweak scale via the 
RGEs, we are able to make predictions for the neutrino masses and mixing angles.  
We obtain lepton mixing angles which are close to the observed values, 
$\theta_{12} =33.8^{\circ}\pm1.2^{\circ}$,  $\theta_{23}=46.9^{\circ}\pm0.9^{\circ}$, and $\theta_{13}=8.56^{\circ}\pm0.20^{\circ}$.
In addition, the neutrino mass-squared differences are found to be $\Delta m^2_{32} = 0.0025\pm0.0001$~eV$^2$ and 
$\Delta m^2_{21}  = 0.000075\pm0.000003$~eV with $m_1=0.0150\pm0.0002$~eV, $m_2=0.0173\pm0.0002$~eV,
and $m_3=0.053\pm 0.002$~eV so that $\sum_i m_i = 0.085\pm0.002$~eV, consistent with experimental observations.

\end{titlepage}

\setcounter{footnote} 0

\newpage

\pagestyle{plain}
\setcounter{page} 1

\section{Introduction}

One of the most significant challenges in high-energy physics 
today is to explain the pattern of masses and mixing angles
exhibited by the elementary fermions in the 
Standard Model (SM).  In the cases of the quarks 
and charged leptons, the masses are strongly hierarchical,
while the masses of the neutrinos are known to be very small in comparison.  
In addition, the quark mixing angles are relatively small while in contrast,
some of the mixing angles for the neutrinos are quite large.  
An explanation for the differences in masses and mixing angles 
between the neutrinos and quarks is currently somewhat of a mystery, 
though the seesaw mechanism does provide a way to obtain naturally
small neutrino masses.  

There has been some progress towards understanding the origin
of the quark and lepton masses and mixing angles by extending the SM
to include  discrete flavour symmetries. 
Indeed, one of the most promising such discrete symmetries is $\mathbf{\Delta(27)}$.
With this discrete symmetry, it has been shown that it is possible to explain the 
masses and mixing angles for quarks and leptons~\cite{Abbas:2014ewa}.  In particular, 
$\mathbf{\Delta(27)}$ contains $\mathbf{A4}$ as a subgroup, and it is known that mass
matrices resulting from imposing an $\mathbf{A4}$ symmetry  naturally leads to 
tribimaximal mixing, which may be taken as a zeroth-order mixing for the
neutrinos. 
Although it might be possible to completely understand the origin 
of quark and lepton mixing by imposing such discrete flavour symmetries, 
this is still somewhat {\it ad-hoc}.  The actual origin of these symmetries remains unexplained.  In fact, ultimately it should be possible to trace the 
origin of these discrete flavour symmetries back to some fundamental theory.
String theory is a leading candidate for such a theory. 

Recently, it has been shown in a particular string model constructed 
in Type IIA string theory with intersecting D-branes that the mass matrices
for the quarks and leptons are the same as those which are obtained
by imposing a $\mathbf{\Delta(27)}$ flavour symmetry~\cite{Mayes:2019isy}. 
Furthermore,
it was demonstrated that it is possible to obtain mass matrices 
for the quarks and charged leptons which results in the correct masses
as well as the correct CKM quark mixing matrix.  In addition, 
it is also possible to simultaneously  obtain a Dirac mass matrix for the neutrinos 
which results in tribimaximal mixing. 
Our approach then was to use the known masses for the quarks and charged leptons as inputs,
as well as the tribimaximal constraint
in order to completely determine the  Dirac neutrino mass matrix.   
  These results are highly non-trivial
as the mass matrices for quarks and leptons in the model are not independent.

Although the results for the neutrinos give the correct mass-squared differences and the
correct mixing matrix, these results are calculated at the string scale, which in the following
we take to be the standard GUT scale,$ M_{GUT}=2\cdot 10^{16}$~GeV. In order to make 
a valid comparison with experimental observations, it is necessary to evolve the neutrino mass 
parameters down to the electroweak scale, $M_{EW}=100$~GeV. 
In the following, we perform a Renormalization Group Equation (RGE) analysis using
the {\tt REAP 11.4 }  Mathematica package~\cite{Antusch:2005gp,Antusch:2003kp}.
We obtain lepton mixing angles which are close to the observed values, 
$\theta_{12}=33.8^{\circ}\pm1.2^{\circ}$, $\theta_{23}=46.9^{\circ}\pm0.9^{\circ}$, and $\theta_{13}=8.56^{\circ}\pm0.20^{\circ}$.
In addition, the neutrino mass-squared differences are found to be $\Delta m^2_{32} = 0.0025\pm0.0001$~eV$^2$ and 
$\Delta m^2_{21} = 0.000075\pm0.000003$~eV$^2$ with $m_1=0.0150\pm0.0002$~eV, $m_2=0.0173\pm0.0002$~eV,
and $m_3=0.053\pm 0.002$~eV so that $\sum_i m_i = 0.085\pm0.002$~eV, consistent with experimental observations.

\section{Neutrino Masses and Mixing Angles}

In recent years, precision measurements of the 
neutrino mixing angles as well as the squares of the mass differences
between neutrinos have been made by several experiments. 
The best estimate of the difference in the squares of the masses 
of mass eigenstates 1 and 2 was published by KamLAND in 2005:
$\Delta m^2_{21}= 0.0000739^{+0.21}_{-0.20}$~eV$^2$~\cite{Araki:2004mb,Esteban:2018azc,Esteban:2016qun,NuFit}.
In addition, the MINOS experiment measured oscillations from an intense 
muon neutrino beam, determining the difference in the squares of the masses 
between neutrino mass eigenstates 2 and 3. Current measurements indicate 
$\Delta m^2_{32}= 0.0027$~eV$^2$~\cite{Esteban:2018azc,Esteban:2016qun,NuFit},
consistent with previous results from Super-Kamiokande~\cite{Fukuda:1998fd}.
In addition, recent analysis of cosmological results constrains the sum of the three neutrino 
masses to be $\lesssim 0.12$~eV~\cite{Vagnozzi:2017ovm}, while additional 
analysis of combined data sets results in $0.15$~eV~\cite{Giusarma:2016phn} 
and $0.19$~eV~\cite{Giusarma:2018jei} for the upper limit. Older analyses set the upper limit
slightly higher at $0.3$~eV~\cite{Thomas:2009ae,Ade:2013zuv,Battye:2013xqa}.

The lepton mixing matrix or PMNS matrix may be parameterized 
as
\begin{equation}
U_{PMNS}   = \left(\begin{array}{ccc}
c_{12}c_{13}& s_{12}c_{13} & s_{13}e^{-i\delta_{CP}}\\
-s_{12}c_{23}-c_{12}s_{23}s_{13}e^{i\delta_{CP}} & c_{12}c_{23}-s_{12}s_{23}s_{13}e^{i\delta_{CP}} &  s_{23}c_{13} \\
s_{12}s_{23}-c_{12}c_{23}s_{13}e^{i\delta_{CP}} & -c_{12}s_{23}-s_{12}c_{23}s_{13}e^{i\delta_{CP}} &  c_{23}c_{13}\end{array} \right),
\label{TBMMassMatrix}
\end{equation}
where $s_{ij}$ and $c_{ij}$ denote $\mbox{sin}~\theta_{ij}$ and $\mbox{cos}~\theta_{ij}$ respectively, 
while ${\delta_{CP}}$ is a $CP$-violating phase.

The current best-fit values for the mixing angles from direct and indirect experiments are,
using normal ordering~\cite{Esteban:2018azc,Esteban:2016qun,NuFit},
\begin{eqnarray}
\theta_{12}=&33.82^{\circ+0.78^{\circ}}_{~-0.76^{\circ}}, \\  \nonumber
\theta_{23}=&49.6^{\circ+1.0^{\circ}}_{~-1.2^{\circ}}\\  \nonumber
\theta_{13}=&~8.61^{\circ+0.13^{\circ}}_{~-0.13^{\circ}}\\  \nonumber
\delta_{CP}=&215^{\circ+40^{\circ}}_{~-29^{\circ}}\\  \nonumber
\end{eqnarray}
 Using these values, the $3\mathbf{\sigma}$ ranges on the PMNS matrix~\cite{Esteban:2018azc} are given by
\begin{equation}
|V|^{3\mathbf{\sigma}}_{PMNS}  = \left(\begin{array}{ccc}
0.797\rightarrow0.842& 0.518\rightarrow0.585 & 0.143\rightarrow0.156 \\
0.233\rightarrow0.495& 0.448\rightarrow0.679&0.639\rightarrow0.783 \\
0.287\rightarrow0.532& 0.486\rightarrow0.706& 0.604\rightarrow0.754
\end{array} \right).
\label{exppmns}
\end{equation}

One of the most studied patterns of neutrino 
mixing angles is the so-called tribimaximal mixing of the form
\begin{equation}
U_{TB}   \sim  \left(\begin{array}{ccc}
\sqrt{\frac{2}{3}} & \sqrt{\frac{1}{3}} & 0\\
-\sqrt{\frac{1}{6}} &  \sqrt{\frac{1}{3}} &  -\sqrt{\frac{1}{2}} \\
- \sqrt{\frac{1}{6}} & \sqrt{\frac{1}{3}}& \sqrt{\frac{1}{2}} \end{array} \right),
\label{TBMMassMatrix}
\end{equation}
which was consistent with early data.  However, the measurement of a 
nonzero $\theta_{13}$ by Data Bay~\cite{An:2012eh} and 
Double Chooz~ \cite{Abe:2011fz}, and confirmed by RENO~\cite{Ahn:2012nd} has 
now ruled out these mixing patterns.  
However, tribimaximal mixing may still be 
viewed as a zeroth-order approximation to more general forms of 
the PMNS matrix which are also consistent with the current data.  
Thus, it is still of great importance to understand the origin of
tribimaximal mixing.  

It has been shown that mass matrices leading to 
tribimaximal and near-tribimaximal mixing may be generated by 
imposing a flavour symmetry such as $\mathbf{A4}$~\cite{Ma:2005qf} 
or $\mathbf{\Delta(27)}$~\cite{Ma:2007wu}.
Specifically, a mass matrix of the form
\begin{equation}
\mathcal{M_{\nu}}   \sim  \left(\begin{array}{ccc}
 Y & X & X \\
X & y & x \\
X & x & y \end{array} \right),
\label{TBMMassMatrix}
\end{equation}
obtained by imposing an $\mathbf{A4}$ flavour symmetry  leads to tribimaximal mixing, 
while mass matrices of the form
\begin{equation}
\mathcal{M_{\nu}}   \sim  \left(\begin{array}{ccc}
f_1 v_1 & f_2 v_3 & f_2 v_2 \\
f_2 v_3 & f_1 v_2& f_2 v_1 \\
f_2 v_2 & f_2 v_1& f_1 v_3\end{array} \right),
\label{NearTBMMassMatrix}
\end{equation}
obtained by imposing an  $\mathbf{\Delta(27)}$ flavour symmetry  may lead to near-tribimaximal mixing.
It is shown in the next section that the Yukawa matrices in a 
particular intersecting D-brane model may naturally be of this form.

In order to naturally explain the smallness of the neutrino masses in comparison
to the other fermion masses, a seesaw mechanism is usually invoked.  
In the canonical or Type I seesaw, the Majorana mass matrix for 
left-handed neutrinos is given by 
\begin{equation}
M^M_{\nu} = - M^D_{\nu}~M^{-1}_R~(M^D_{\nu})^T,
\label{seesaw}
\end{equation}
where $M^D_{\nu}$ is the Dirac mass matrix for neutrinos 
and $M_R$ is the right-handed neutrino mass matrix.

\section{Fermion Mass Matrices in a Realistic String Model }

 In the past two decades, a
promising approach to string model building has emerged involving
compactifications with D-branes on orientifolds (for reviews,
see~\cite{Ura03,BluCveLanShi05, BluKorLusSti06,Mar07}).  In such
models chiral fermions---an intrinsic feature of the Standard Model
(SM)---arise from configurations with D-branes located at transversal
orbifold/conifold singularities~\cite{DouMoo96} and strings stretching
between D-branes intersecting at
angles~\cite{BerDouLei96,AldFraIbaRabUra00} (or, in its T-dual
picture, with magnetized D-branes~\cite{Bac95,BluGorKorLus00,AngAntDudSag00}).

 Within the framework of
D-brane modeling it was demonstrated that the Yukawa matrices $Y_{abc}
\sim \exp(- A_{abc})$ arise from worldsheet areas $A_{abc}$ spanning D
branes (labeled by $a$, $b$, $c$) supporting fermions and Higgses at
their intersections~\cite{AldFraIbaRabUra00,Cremades:2003qj}.  This
pattern naturally encodes the hierarchy of Yukawa couplings. In addition,
due to the internal geometry of these compactifications as well as due to 
stringy selection rules present in such models, discrete flavour symmetries
may arise.  In particular, it has been shown that discrete symmetries such 
as $D_4$ and $\Delta(27)$ may naturally arise in such models~\cite{Abe:2009vi}.  

 However,
for most string constructions, the Yukawa matrices are of rank one.  In
the case of D-brane models built on toroidal orientifolds, this result
can be traced to the fact that not all of the intersections at which
the SM fermions are localized occur on the same torus.  To date only
one three-generation model is known in which this problem has been
overcome~\cite{Chen:2007px,Chen:2007zu}, and for which one can obtain mass
matrices for quarks and leptons that may reproduce the experimentally observed
values.  Additionally, this model exhibits automatic gauge coupling
unification at the string scale, and all extra matter can be
decoupled.  

In this model, 
the Yukawa couplings for the quarks and leptons are all allowed and are given by the superpotential
\begin{equation}
W_Y = Y^U_{ijk} Q_L^i U_R^j H_U^k + Y^D_{ijk} Q_L^i D_R^j H_D^k + Y^{\nu}_{ijk} L^i N^j H_U^k + Y^L_{ijk} L^i E^j H_D^k, 
\end{equation}
where  $Q_L^ii$ and $L^i$ are the left-handed quark and lepton fields respectively, while $U_R^j$, $D_R^j$, $N^j$, and $E^j$ are the
right-handed up quarks, down quarks, neutrinos, and charged leptons respectively, and $H_U^k$ and $H_D^k$ are the up-type and down-type 
Higgs fields, with
\begin{eqnarray}
i \in \left\{0, 1, 2 \right\}, \ \ \ \ \ j \in
\left\{0, 1, 2\right\}, \ \ \ \ \ k \in
\left\{0, 1, 2, 3, 4, 5\right\}.
\end{eqnarray}
In addition the $\mu$ term and right-handed neutrino masses
which may be generated via the following  higher-dimensional
operators:
\begin{eqnarray}
W \supset &&{{y^{ijkl}_{\mu}} \over {M_{\rm St}}} S_L^i S_R^j
H_u^k H_d^l + {{y^{mnkl}_{Nij}}\over {M^3_{\rm St}}} T_R^{m}
T_R^{n} \Phi_i \Phi_j  F_R^k  F_R^l ~,~\,
\label{eqn:HiggsSup}
\end{eqnarray}
where $S_L^i$, $S_R^j$, $\Phi_j$,  $T_R^m$ anre $SU(2)_L$ and $SU(2)_R$  singlet and triplet
fields respectively present in the model~\cite{Chen:2007zu}.   

A complete form for the Yukawa couplings $y^f_{ij}$ for D6-branes
wrapping on a full compact space $T^2 \times T^2 \times T^2$ can be
expressed as~\cite{CvePap06,Cremades:2003qj}:
\begin{equation} \label{Yukawas}
Y_{\{ijk\}}=h_{qu} \sigma_{abc} \prod_{r=1}^3 \vartheta
\left[\begin{array}{c} \delta^{(r)}\\ \phi^{(r)}
\end{array} \right] (\kappa^{(r)}),
\end{equation}
where
\begin{equation}
\vartheta \left[\begin{array}{c} \delta^{(r)}\\ \phi^{(r)}
\end{array} \right] (\kappa^{(r)})=\sum_{l \in\mathbf{Z}} e^{\pi
i(\delta^{(r)}+l )^2 \kappa^{(r)}} e^{2\pi i(\delta^{(r)}+l )
\phi^{(r)}},   \label{Dtheta}
\end{equation}
with $r=1,2,3$ denoting the three two-tori. For the present model, we focus 
on the first torus ($r=1$) as the other two-tori only produce
and overall constant.

The parameter $\delta^{(1)}$ is a function of $i$, $j$, and $k$ 
and is given by 
\begin{equation}
\delta^{(1)} = \frac{i}{3}-\frac{j}{3}-\frac{k}{6} + \frac{s}{3} + \epsilon^{(1)}_{(a,b)},
\end{equation}
 By choosing a different linear function for $s^{(1)}$, some independent
modes with non-zero eigenvalues are possible. Specifically, we 
will consider the case 
$s^{(1)}= -i$ so that 
\begin{equation}
\delta^{(1)} = -\frac{j}{3}-\frac{k}{6} + \epsilon^{(1)}_{(a,b)},
\end{equation}
Here, the parameter $\epsilon{(1)}_{a,bl}$ is an overall shift
parameter which depends upon the positions of 
each stack of D-branes in the internal space. Thus for a quark or lepton field localized
at the intersection between stacks $a$ and $b$, the shift parameter $\epsilon_{(a,b)}$ depends
on the positions of stacks $a$ and $b$ in the internal space.  In 
order to have a consistent solution, the shift parameters
for each type of fermion must satisfy the constraint,
\begin{equation}
\epsilon_u^{(r)} + \epsilon_{l}^{(r)} = \epsilon_d^{(r)} + \epsilon_{\nu}^{(r)}.
\end{equation}  
 
Likewise, the parameter $\kappa^{(1)}=\frac{6J^{(1)}}{\alpha'}$ is a scale-factor related to the  K\a"ahler
modulus  $J^{(1)}$, while $\phi^{(r)}$ is an effective Wilson 
line for each torus. See~\cite{Mayes:2019isy}
for the full definition of these parameters.  

In addition, 
there is a selection rule,
\begin{equation}
i + j + k = 0 \ \mbox{mod} \ 3. 
\label{selrule} 
\end{equation}
which determines which Higgs fields couple to the different 
quark and lepton fields.  
The
Yukawa matrices in this model  are then 
of the form
\begin{equation}
\mathcal{M}   \sim  \left(\begin{array}{ccc}
A v_1 & B v_3  &  C v_5 \\
C v_3 & A v_5 & B v_1 \\
B v_5 & C v_1 & A v_3 \end{array} \right)
 +
\left(\begin{array}{ccc}
E v_4 &  F v_6 & D v_2 \\
D v_6 &  E v_2 & F v_4 \\
F v_2&D v_4 & E v_6 \end{array} \right), 
\label{Yukawa general2}
\end{equation}	
where $v_k = \left\langle H_{k+1} \right\rangle$ and  
the Yukawa couplings 
$A$, $B$, $C$, $D$, $E$, and $F$ are given by
\begin{eqnarray}
&&A \equiv\vartheta \left[\begin{array}{c}
\epsilon^{(1)}\\ \phi^{(1)} \end{array} \right]
(\frac{6J^{(1)}}{\alpha'}),  \ \ \ \ \ \ \ \ \ \ \ \ \ \ \ 
B \equiv \vartheta \left[\begin{array}{c}
\epsilon^{(1)}+\frac{1}{3}\\  \phi^{(1)} \end{array} \right]
(\frac{6J^{(1)}}{\alpha'}),  \nonumber \\
&&C \equiv \vartheta \left[\begin{array}{c}
\epsilon^{(1)}-\frac{1}{3}\\  \phi^{(1)} \end{array} \right]
(\frac{6J^{(1)}}{\alpha'}), \ \ \ \ \ \ \ \ \ \ 
D \equiv \vartheta \left[\begin{array}{c}
\epsilon^{(1)}+\frac{1}{6}\\  \phi^{(1)} \end{array} \right]
(\frac{6J^{(1)}}{\alpha'}), \nonumber \\
&&E \equiv \vartheta \left[\begin{array}{c}
\epsilon^{(1)}+\frac{1}{2}\\  \phi^{(1)} \end{array} \right]
(\frac{6J^{(1)}}{\alpha'}), \ \ \ \ \ \ \ \ \ \ 
F \equiv \vartheta \left[\begin{array}{c}
\epsilon^{(1)}-\frac{1}{6}\\  \phi^{(1)} \end{array} \right]
(\frac{6J^{(1)}}{\alpha'}).
\end{eqnarray}  

These Yukawa
matrices are of rank 3, such that it is possible to have three 
different mass eigenvalues as well as non-trivial  mixing between each of the different 
generations.  In the MSSM, the up-type quarks and neutrinos receive mass from the
isospin up  Higgs field $H_U$, while the down-type quarks and charged leptons receive mass 
from the isospin down Higgs field $H_D$.  In this model, there are actually six different 
Higgs fields in each sector.  We may fine-tune the Higgsino bilinear$\mu$-term given in 
Eq.~\ref{eqn:HiggsSup}  such that
there are only two massless eigenstates given by
\begin{equation}
H_{u,d} = \sum_i \frac{v^i_{u,d}}{\sqrt{\sum(v^i_{u,d})^2}},
\label{muterm}
\end{equation}
where $v^i_{u,d} = \left\langle H^i_{u,d} \right\rangle$.
In fitting the mass matrices, we treat the Higgs VEVs as free parameters.
However, these parameters may ultimately be calculated in the model.  
By choosing the shift parameter to be $\epsilon = 0$ for quarks 
and $\epsilon = 1/2$ for leptons (or {\it vice-versa}), the mass matrices are of the same form as those
obtained by imposing a $\mathbf{\Delta(27)}$ discrete flavour symmetry given in Eq.~\ref{NearTBMMassMatrix}, 
since for these values of the shift parameters, $B=C$ and $D=F$. 
In addition with this choice, 
the up and down-type quarks
predominantly receive masses mainly via the odd-numbered Higgs VEVs 
$v^{U,D}_{\mbox{odd}}$ while the neutrinos and charged leptons obtain
mass predominately via the even-numbered Higgs VEVs $v^{U,D}_{\mbox{even}}$, 
or {\it vice-versa}. However, it should be emphasized that the fermions in each sector
couple to all of the Higgs fields in each sector. Thus, the mass matrix for the up-type quarks 
is not independent from the neutrino mass matrix, and likewise for the down-type quarks 
and the charged leptons. 

 Our strategy then is to choose the Higgs VEVs 
and the K\a"ahler parameter $\kappa$ in order to obtain mass matrices
for the quarks and leptons that give the correct mass eigenvalues as 
well as the correct CKM matrix. In particular, we fit them for tan~$\beta=50$ as shown in~\cite{Mayes:2019isy}.
This may be accomplished  by choosing the following values for 
the Higgs VEVs:
\begin{equation}
\begin{array}{l c l}
 v^1_u= 0.0000142, && v^1_d=0.0028224 \\
v^2_u= 0.00002408185, && v^2_d=0.045    \\
v^3_u= 1.0, && v^3_d=1.0     \\
v^4_u= 0.000000345,&& v^4_d=0.0010105      \\
v^5_u=0.00404, && v^5_d=0.0266   \\
v^6_u= 0.005960855, && v^6_d=1.0.
\end{array}
\end{equation}
Nearly the correct CKM matrix is then  obtained by choosing $\kappa=58.7$.  
Note that the Higgs VEVs $v^{even}_u$ have been chosen so that the neutrino mass matrix will be near-tribimaximal,
i.e. in the form given in Eq.~\ref{TBMMassMatrix}.
The values for  VEVs $v^{even}$  required to do this are then determined by 
the off-diagonal elements of the up-type quark mass matrix when those Yukawa couplings are evaluated at $\epsilon=0.5$.
Thus, the up-type quark mass matrix and the Dirac neutrino mass matrix are not independent of each other. 
Once the up-type quark matrix is determined, the requirement that the neutrino matrix result in tribimaximal mixing completely fixes it. 
For example, after fixing the odd-numbered up-type Higgs VEVs, the neutrino mass matrix with $\epsilon_{\nu}=0.5$ is given by 
\begin{equation}
M_{\nu} = m_t \left(\begin{array}{ccc}
Ev^4_u & 0.005960855+Fv^6_u & 0.0002408185+Dv^2_u  \\
 0.005960855+Dv^6_u & 0.00005+Ev^2_u &0.00000008464414+Fv^4_u \\
0.0002408185+Fv^2_u &0.00000008464414+Dv^4_u & Ev^6_u
\end{array} \right). \;\; 
\label{umass}
\end{equation}
Then $v^2_u$, $v^4_u$, $v^6_u$ may be chosen so that the neutrino mass matrix  is of the form given in   Eq.~\ref{TBMMassMatrix}
which results in tribimaximal mixing.  

We set all $CP$ phases to zero by setting the Wilson lines, $\phi$  which are input into the Jacobi Theta functions equal to zero.  However, 
note that they may also be included in the fit so that the Dirac CP violating  phase, which appears in the neutrino mixing matrix, may be determined once 
the CP violating phases in the quark sector are fit.  
With these parameters, the mass matrices for the up
and down-type quarks are given by
\begin{equation}
M_u = m_t \left(\begin{array}{ccc}
0.0000142& 0.00003553304 & 0.0000001436  \\
 0.00003553304 & 0.00404 & 0.000000002055 \\
0.0000001436 & 0.000000002055 & 1
\end{array} \right), \;\; 
\label{umass}
\end{equation}
\begin{equation}
M_d = m_b \left(\begin{array}{ccc}
0.0028224 & 0.005960856 & 0.0002682385  \\
 0.005960856 & 0.0266 & 0.000006023448 \\
0.0002682385 & 0.000006023448 & 1
\end{array} \right),
\label{dmass}
\end{equation}
whose eigenvalues have the correct quark mass hierarchies and nearly 
the correct CKM matrix is obtained.  
Similarly, the mass matrices for the 
neutrinos and charged leptons are given by
\begin{equation}
M_{\nu} = m_t \left(\begin{array}{ccc}
0.000000345& 0.005960855 & 0.00002408185  \\
 0.005960855 & 0.00002408185 & 0.00000008464414\\
0.00002408185 &  0.00000008464414 & 0.005960855
\end{array} \right), \;\; 
\label{numass}
\end{equation}
\begin{equation}
M_l = m_b \left(\begin{array}{ccc}
0.0010105 & 0.005960856 & 0.0001585588 \\
0.005960856 & 0.045 & 0.00001682392 \\
 0.0001585588 & 0.00001682392 & 1
\end{array} \right).
\label{lepmass}
\end{equation}
The eigenvalues for the charged lepton mass matrix then have the correct mass hierarchy, 
while the Dirac neutrino mass matrix is near-tribimaximal
and  the charged lepton mass matrix is near-diagonal.  

\section{RGE Evolution to the Electroweak Scale}

The Yukawa mass matrices obtained in the previous section are 
so obtained at the string scale, which in the following we will
take to be the same as the GUT scale, $M_X = 2\cdot10^{16}$~GeV.
In order to compare the model predictions with experimental 
results it is necessary to evolve these mass matrices down 
to the electroweak scale via the Renormalization Group Equations (RGE),
as well as apply a seesaw mechanism.  In order to do this we use the 
{\tt REAP 11.4 } Mathematica package~\cite{Antusch:2003kp}.

\begin{figure}
	\centering
		\includegraphics[width=0.5\textwidth]{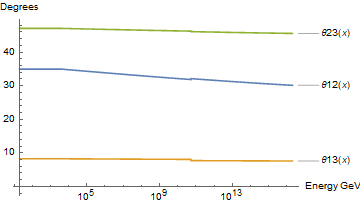}
		\caption{Neutrino mixing angles as a function of energy scale. Note that we have taken the supersymmetry decoupling scale to be 
$2-7$~TeV in order to obtain the best agreement with data.  }
	\label{fig:MixingAngles}
\end{figure}

\begin{figure}
	\centering
		\includegraphics[width=0.5\textwidth]{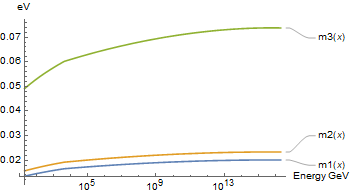}
		\caption{Neutrino masses as a function of energy scale. Note that we have taken the supersymmetry decoupling scale to be 
$2-7$~TeV in order to obtain the best agreement with data.  }
	\label{fig:NeutrinoMasses}
\end{figure}

\begin{figure}
	\centering
		\includegraphics[width=0.5\textwidth]{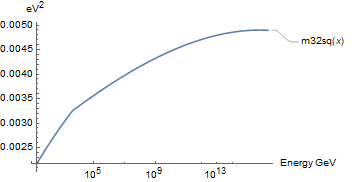}
		\caption{Mass difference $m^2_{32}$ as a function of energy scale. Note that we have taken the supersymmetry decoupling scale to be 
$2-7$~TeV in order to obtain the best agreement with data. }
	\label{fig:m32sq}
\end{figure}

\begin{figure}
	\centering
		\includegraphics[width=0.5\textwidth]{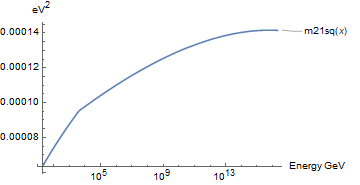}
		\caption{Mass difference $m^2_{32}$ as a function of energy scale. Note that we have taken the supersymmetry decoupling scale to be 
$2-7$~TeV in order to obtain the best agreement with data. }
	\label{fig:m21sq}
\end{figure}

The Yukawa couplings are given by
\begin{eqnarray}
y_t = \frac{\sqrt{2}m_t}{v\cdot\mbox{sin}\beta}\,  \ \ \ \ \ \ \ \ \ \ \ \ \ \ \  \ \ \ \ \ \ \ \ \ y_b=\frac{\sqrt{2}m_b}{v\cdot\mbox{cos}\beta},\ \ \ \ \ \ \ \ \ \  
\end{eqnarray}
where $m_t = 129$~GeV and $m_b=1.00$~GeV at $M_{GUT}$. For $\mbox{tan}\beta=50$, 
we then have that $y_t = 0.742$ and $y_b=0.286$. The Yukawa matrices are then given by
\begin{equation}
Y_u = y_t \left(\begin{array}{ccc}
0.0000142& 0.00003553304 & 0.0000001436  \\
 0.00003553304 & 0.00404 & 0.000000002055 \\
0.0000001436 & 0.000000002055 & 1
\end{array} \right), \;\; 
\label{yuk}
\end{equation}
\begin{equation}
Y_d = y_b \left(\begin{array}{ccc}
0.0028224 & 0.005960856 & 0.0002682385  \\
 0.005960856 & 0.0266 & 0.000006023448 \\
0.0002682385 & 0.000006023448 & 1
\end{array} \right),
\label{dyuk}
\end{equation}
and 
\begin{equation}
Y_{\nu} = y_t \left(\begin{array}{ccc}
0.000000345& 0.005960855 & 0.00002408185  \\
 0.005960855 & 0.00002408185 & 0.00000008464414\\
0.00002408185 &  0.00000008464414 & 0.005960855
\end{array} \right), \;\; 
\label{nyuk}
\end{equation}
\begin{equation}
Y_l = y_b \left(\begin{array}{ccc}
0.0010105 & 0.005960856 & 0.0001585588 \\
0.005960856 & 0.045 & 0.00001682392 \\
 0.0001585588 & 0.00001682392 & 1
\end{array} \right).
\label{lepyuk}
\end{equation}

In addition to the Yukawa matrices for quarks and leptons, we must choose a right-handed neutrino mass matrix
to  be input into the seesaw mechanism, resulting in the mass matrix for left-handed neutrinos given by Eq.~\ref{seesaw}. 
 In principle, this may be calculated in the model.  However, the right-handed
neutrino mass term  in the superpotential arises from dimension-five operators as shown in Eq,~\ref{eqn:HiggsSup}, 
thus it is very difficult to calculate.  
Therefore, for the present work we will choose the right-handed neutrino mass matrix such that the near-tribimaximal 
neutrino mixing originating in the Dirac neutrino mass matrix is preserved when running the RGEs from the string scale
down to the GUT scale.  A scan over a large number of random matrices produced twenty-thousand different choices 
for the right-handed neutrino matrix.  These right-handed neutrino matrices produced neutrino masses which give mass-squared differences within the
experimentally observed ranges with a ratio of $\Delta m^2_{32} / \Delta m^2_{21} > 30$.  These right-handed neutrino matrices also produced neutrino masses whose sum is lower than the upper limit from
cosmological data defined as $\sum_i m_i \lesssim 0.12$~eV, and results in a PMNS matrix within the $3\sigma$ ranges given in
 Eq.~\ref{exppmns}. The values for the right-handed neutrino matrix elements which satisfy these constraints are

\begin{equation}
M_R \approx M_r\left(\begin{array}{cccc}
-5.01\pm0.05 & -0.24\pm0.01& 0.25\pm0.01 \\
-0.24\pm0.01 & -3.12\pm0.04& 1.69\pm0.07 \\
  0.25\pm0.01& 1.69\pm0.07 & -3.22\pm0.07
\end{array} \right)
\label{RHMass}
\end{equation}
where $M_r=10^{10}$~GeV\footnote{It may also be possible to employ an inverse seesaw mechanism using lower dimensional operators which are easier to calculate}.  
Inserting these values into the {\tt REAP} package, we then find that the neutrino mixing angles at the electroweak scale are given by
\begin{eqnarray}
\theta_{12} = 33.8^{\circ}\pm 1.2^{\circ}, \ \ \ \ \ \ \ \ \ \ 
\theta_{23} = 46.9^{\circ}\pm 0.9^{\circ}, \ \ \ \ \ \ \ \ \ \ 
\theta_{13} = 8.56^{\circ}\pm 0.20^{\circ}, 
\end{eqnarray}
where we give an uncertainty based upon how these results change when the right-handed neutrino mass
parameters are varied.  
Using these values for the mixing angles, the PMNS lepton mixing matrix is then given by
\begin{equation}
|V_{PMNS}| \approx \left( \begin{array}{ccc}
0.8 22\pm0.01&  0.549\pm0.01& 0.149\pm0.003\\
0.470\pm0.012& 0.5081\pm0.016& 0.721\pm0.010 \\
0.321\pm0.017& 0.663\pm9.011 & 0.6676\pm0.011
\end{array} \right),
\end{equation}
which is in excellent agreement with the $3\sigma$  limits given in 
Eq.~\ref{exppmns}.
A plot of the neutrino mixing angles as a function of energy scale is 
shown in Fig.~\ref{fig:MixingAngles}, while the running of the neutrino masses
is shown in Fig.~\ref{fig:NeutrinoMasses}.

In addition, we find that the neutrino masses at the electroweak scale are given by 
\begin{eqnarray}
m_1 = 0.0150\pm0.0002~\mbox{eV}, \ \ \ \ \ \ \ \ \ \ 
m_2 = 0.0173\pm0.0002~\mbox{eV}, \ \ \ \ \ \ \ \ \ \  
m_3 = 0.053\pm0.002~\mbox{eV}, \ \ \ \ \ \ \ \ \ \ 
\end{eqnarray}
with 
\begin{equation}
\sum_i m_i = 0.085\pm0.002~\mbox{eV},
\end{equation}
consistent with cosmological constraints.  
Finally, we find that 
\begin{eqnarray}
\Delta m^2_{32} = 0.0025\pm0.0002~\mbox{eV}^2, \ \ \ \ \ \ \ \ \ \  \ \ \ \ \ 
\Delta m^2_{21} = 0.000075\pm0.000003~\mbox{eV}^2.
\end{eqnarray}
These values are consistent with current experimental observations of neutrino oscillations.
Plots of $m^2_{32}$
and $m^2_{21}$ as functions of the energy scale are shown in 
Fig.~\ref{fig:m32sq} and Fig.~\ref{fig:m21sq}. 
Note that we have taken the supersymmetry decoupling scale to be 
$2-7$~TeV in order to obtain the best agreement with data.  
Thus, a change of slope in the RGE plots may be seen 
beginning at this scale.

\section{Conclusion}

We have performed an RGE analysis of the neutrino masses and mixing angles 
in a realistic  Pati-Salam model constructed from intersecting D6 branes 
on a $T^6/(\Z_2 \times \Z_2)$ orientifold. In previous work it had been shown  that it is possible to fit the quark and lepton Yukawa matrices in the model
such that the correct masses are obtained for the quarks and charged leptons as well as the nearly correct CKM quark mixing matrix. In addition, a Dirac
neutrino mass matrix which is nearly tribimaximal was naturally obtained. A suitable right-handed neutrino mass matrix was chosen and then inserted into a Type I seesaw mechanism
along with the Dirac neutrino mass matrix. 
The neutrino mass parameters were then evolved from the GUT scale down to the electroweak scale using the {\tt REAP} Mathematica package.  
We then obtained neutrino masses given by $m_1=0.015\pm0.0002$~eV, $m_2=0.0173\pm0.0002$~eV, and $m_3=0.053\pm0.002$~eV, while simultaneously 
obtaining electroweak scale mixing angles and neutrino mass-squared differences which are within current experimental limits.

In fitting the SM fermion masses and mixings we have made use of several  free parameters.  In particular, the free parameters are 
the twelve Higgs VEVs, the K\a"ahler parameter $\kappa$, and the five independent parameters in the right-handed neutrino mass matrix.  
In addition there are four shift parameters which are fixed by the $\mathbf{\Delta(27)}$ constraint on the mass matrices.  In addition, there
is the effective supersymmetry decoupling scale in the RGE analysis which we have fixed by requiring the neutrino mass-squared differences to be consistent with
current experimental data.  Therefore, there is nominally a total of eighteen free parameters in the analysis.  However, since these parameters are not independent of each other
there is effectively  less than eighteen.  
With these parameters we have fit twelve quark and lepton masses and six quark and lepton mixing angles for a total of eighteen observable quantities.  
Therefore, there are no more  free parameters than there are observable quantities.  Thus, as the fit is highly constrained, the obtained values for the neutrino masses
may be regarded as a bona-fide prediction of the model.  

In principle, it may be possible to determine all of the adjustable parameters within the model.  For example, 
by calculating the Higgsino bilinear mass matrix given in Eq.~\ref{muterm} it may be possible to determine
the values of the Higgs VEVs.  In particular, they would correspond to the coefficients for the massless eigenstates 
corresponding to $H_U$ and $H_D$.  Similarly, the right-handed neutrino mass matrix may in principle be calculated 
within the model.  Then, it might be possible to calculate the observed masses and mixing angles for the quarks and leptons
from first principles. Another possibility is that the observed CP-violating phases appearing in the CKM matrix may be included in
the fit, allowing the Dirac CP phase appearing in the lepton mixing matrix to be predicted. 
We plan to explore these possibilities in future work.

\paragraph{Acknowledgments.} Evan Howington was supported by the grant, Pathways to STEM Careers, funded by the HSI STEM program of DOE. 

%


\end{document}